\documentclass[twocolumn]{revtex4}
\usepackage{epsfig}

\usepackage{amsmath}
\begin{document}
\title{ The cosmological behaviour and the statefinder diagnosis for the New Tsallis agegraphic dark energy }
\author{Umesh  Kumar Sharma$^{1}$\footnote{sharma.umesh@gla.ac.in},  Shikha Srivastava$^{2}$\footnote{shikha.azm06@gmail.com}}
\address{$^{1,2}$Department of Mathematics, Institute of Applied Sciences and Humanities, GLA University\\
	Mathura-281 406, Uttar Pradesh, India}

\begin{abstract}

 In this work, we have considered the recently proposed new Tsallis  Agegraphic Dark Energy model (NTADE) (Mod. Phys. Lett. A 34, 1950086, 2019) within the framework  of a flat Friedmann-Robertson-Walker(FRW)
 Universe by taking various values of the parameter $\delta$.  The NTADE model shows the current phase transition of the Universe from
 decelerated to accelerated phase. The NTADE EoS parameter shows a rich behaviour as it can be quintessence-like or phantom-like depending on the value of $\delta$. For discriminating the NTADE model from $\Lambda$CDM, we have plotted the statefinder parameters $r(z)$, $s(z)$ and $(r, s)$, $(r, q)$ pair. The NTADE model shows distinct evolutionary trajectories of their evolution in  ($ r, s$) and ($ r, q$) plane.  An analysis using the snap parameter and
 the $\omega_{D}-\omega_{D}^{'}$ pair dynamical analysis have also been performed.

\end{abstract}
\maketitle

\section{Introduction}

According to a sequence of past and latest observational data \cite{ref1,ref2,ref3,ref4,ref5,ref6}, the dark sector of our Universe is filled  with two dark fluids, namely the dark energy (DE)
 and dark matter (DM). The former fluid drives the current accelerating stage of the Universe while the later i.e. DM is responsible for the structure formation of the Universe. It is also observed from the observational data that about ninety six percent of the total energy density of the Universe is coming from this combine dark component where in particular, the contribution of DE is around sixty eight percent of the total energy allocation of the Universe while the DM contributes approximately twenty eight percent of the  total energy allocation of the Universe. Although, the origin, evolution and the characteristics of these dark sector of the DE are not distinctly recognized yet. However, the nature of DM appears to be partially known by indirect gravitational effects, although, the DE has endured  being exceptionally mysterious. As a consequence, in the last couple of years, various cosmological models have been proposed and
explored.  The non-interacting models of the Universe are considered to be the simplest model leading to two independent evolution of these dark  components where DE and DM are conserved separately. While a more generalized form of the cosmological models are available in which DE and DM are permitted to interact with each other \cite{ref6a,ref6b,ref6c,ref6d,ref6e,ref6f}.\\

To  describe  the  accelerated expansion  of  the  cosmos  motivated  by  holographic  principle \cite{ref23,ref24,ref25,ref26},  M. Li suggested holographic dark energy (HDE) where IR cutoff was taken care by future event horizon  \cite{ref22}.   After  that  Agegrapic  dark  energy  (ADE) model was suggested by Cai by taking length measure as the age of the Universe \cite{ref22a}. Furthermore,  by considering conformal time  as a time  scale,  Wei  and  Cai  suggested  the  New agegraphic dark energy (NADE) model \cite{ref82}. ADE models gained  a lot of interest after the proposal and its cosmological consequences were investigated with observational  constraints \cite{refa1,refa2}.
  Different entropies  are also used  for the investigation  of  the  cosmological  and  gravitational   scenario.  Firstly, it has been shown that differences between Tsallis and Bekenstein entropies can describe DE, modify Friedmann equations, and even make a bridge between Verlinde and Padmanabhan approaches of emergent gravity \cite{refa10}. Recently, three  new form of dark energy models namely, the Tsallis holographic dark energy (THDE) \cite {ref31}, the R$\acute{e}$nyi holographic dark energy (RHDE) \cite {ref32} and the Sharma- Mittal  holographic dark energy (SMHDE) \cite {ref33}  are proposed based on holographic principle and  generalized entropy formalism \cite{ref31a,ref31b,ref31c, ref31d,ref31e,ref31f,ref31g}. These models of  holographic dark energy can be used to clarify or explain the cosmic acceleration of the universe \cite{ref34,ref35,ref37,ref38,ref40,ref42,ref44,ref45,ref47,ref48}.  The Friedmann equations has been  derived  by using  the Tsallis entropy considering the apparent horizon as IR cut off 
  	in FRW Universe, to describe the Universe dynamics \cite{refG1}. Furthermore, to explain the 
  	  	evolution of the FRW Universe,  the Friedmann equations get modified  for large-scale gravitational systems using the non-extensive 
  	Tsallis entropy on the relativistic cosmology background \cite{refG2}.\\

Recently, the authors  proposed a new DE model called Tsallis agegraphic dark energy (TADE) with time scale as IR cut-off and new Tsallis agegraphic dark energy (NTADE) with the conformal time as system IR cut-offs \cite{Zadeh:2018wub}. The cosmological parameters such as the deceleration parameter $q$, the energy density parameter $\Omega_{{D}}$, the equation of state parameter (EoS) $\omega_{{D}}$ and squared sound speed $v_{s}^{2}$ are investigated to study the evolution of the Universe filled with the TADE and the pressure-less DM. in  \cite{Zadeh:2018wub}. They proposed that the late time acceleration of the Universe can be observed by the TADE and NTADE models with and without interaction. Now, modified 
	gravity also became a necessary component of theoretical cosmology. To explain the qualitative transformation 
	of the gravitational interaction in the very early and/or very late Universe, it is aimed as a generalization of general relativity. 
	Appropriately, nowadays modified gravity may not only represent the late-time acceleration and early-time inflation but also introduce 
	the unified consistent classification of the Universe evolution eras series: inflation, matter/ radiation, and DE dominance (For a more recent review see \cite{ref38c1}). In this viewpoint, recently, Zadeh investigated the cosmological consequences  of the  NTADE models with and without interaction in some  of the  modified gravity for example,   modified Chern-Simons gravity \cite{refs1}, Horava-Lifshitz cosmology \cite{refs2}, and  flat Fractal Cosmology \cite{refs3}.\\
	
	 As discussed above, the DE phenomenon can be analyzed by a large number of models. Hence, it is very important to differentiate among these DE models. To achieve this, statefinder pair $(r; s)$ was
	proposed in \cite{Sahni:2002fz,Alam:2003sc}, which is a notable geometrical indicative to
	eliminate the degeneracy in the present value of Hubble parameter $H$ and the present value of the deceleration parameter $q$ of
	different dark energy models. Since $(r; s)$ plane has distinct evolutionary trajectories for different dark energy models. Hence, it is used extensively by researchers to discriminate among
	many modified theories of gravity and various models of dark energy in the literature. The statefinder pair $(r, s)$ \cite{Sahni:2002fz,Alam:2003sc}, are defined  as 
\begin{eqnarray}
\label{eq1}
r = \frac{\dddot {a}}{aH^{3}} ; \hspace{1cm} s = \frac{r-1}{3 (q - \frac{1}{2})},
\end{eqnarray}
where $a$, $H$ and $q$, represent the scale factor, Hubble parameter and deceleration parameter, respectively.\\

 After the proposal of statefinder diagnostic, firstly, the HDE models were discriminated  from $\Lambda$CDM  by Zhang \cite{ref56} through this diagnostic for the  different best fit values of constant $c$, which plays a very significant role in HDE models. In \cite{ref56a}, Wei and Cai explored  the ADE models with and without  interaction through statefinder diagnostic and $\omega_{D} - \omega_{D}^{'}$ pair and proposed that the ADE models can easily be  discriminated from the $\Lambda$CDM model. Gao et al. proposed the Ricci dark energy (RDE) \cite{ref56b} model and it is  explored through statefinder diagnostic by considering ricci scalar as  Universe horizon in a flat FRW Universe \cite{ref54}. In \cite{ref55}, the authors used this geometrical tool to make a classification and a discrimination
about the modified polytropic Cardassian models.\\

   In the DGP braneworld as well as the HDE in standard cosmology, the evolution of the deceleration parameter, EoS parameter and the statefinder
parameters have been investigated considering both GO and Hubble horizon as IR cutoff in a flat FRW Universe \cite{refN1}. For both proposed cutoff, they found that depending on the model parameters, the statefinder pair parameters $(r, s)$ become constant  and  restored those
of $LCDM$ model for the appropriate choice of the model parameters.  The statefinder diagnostic  has been applied in Palatini formalism on a series of $f(R)$ gravity models in a flat FRW Universe for series of the chosen model parameters  to see if they distinguish from one another \cite{refN2}. The statefinder diagnostic is a useful method for discriminating various DE models.  The cosmological evolution and the geometrical behavior using  the statefinder diagnostic  of  the  interacting NADE model has been investigated in a flat FRW Universe \cite{refN3,refN4}. The polytropic gas DE models  have been explored through the statefinder diagnostic and  $\{\omega_D, \omega_D^{'}\}$ pair  in a
flat FRW Universe and proposed that these  models mimic the standard $LCDM$ model at the early time \cite{refN5}.

 Recently, the THDE, RHDE and SMHDE models are investigated with and without interaction between DE and DM from the statefinder viewpoint  for different values of the  non-extensive parameter $\delta$ \cite {ref57,ref58a,ref58b}. A direct comparison has also been performed for these recently proposed DE models through statefinder diagnostic \cite{ref58c}. The statefinder parameters of the different DE models in non-flat FRW Universe have  also been examined with and without interaction in \cite{ref59,ref59a,ref59b,ref60a,ref60b}. The statefinder parameter diagnostic is also used to  discriminate various dark energy models such as the RDE, the new HDE,  the new ADE and the standard HDE model \cite{ref62}.\\

 Let us stress the  similarity and difference of this work with other works.
  	  More recently, the TADE  models are investigated considering the statefinder diagnostic and the  $\{\omega_D, \omega_D^{'}\}$ pair with and without interaction in a flat FLRW Universe \cite{ref61, Xu:2019hhs}, while they have not considered the new Tsallis agegraphic dark energy (NTADE)  models. Therefore, in the present contribution, we shall investigate the cosmological behavior of  the non-interacting new Tsallis agegraphic dark energy (NTADE)
 cosmological model and discriminate the NTADE model through the statefinder diagnostic and the  $\{\omega_D, \omega_D^{'}\}$ pair.
Also,  the evolutionary trajectories the statefinder parameters   are plotted by taking the  NTADE energy density value ($\Omega_{D}^{0}$), ($\Omega_{D}^{0} =0.70$), in consideration of Planck 2018 results VI- $\Lambda$CDM cosmology \cite{ref5}.\\

The rest of the paper is organized as: the  New Tsallis agegraphic Dark Energy  and   cosmological parameters such as deceleration parameter, the NTADE EoS parameter are discussed in Sec. $II$.  To discuss the geometrical behavior of our model, we obtain statefinder parameters in Sec. $III$. Analysis of the $\omega_{D}-\omega_{D}^{'}$ pair has been discussed in Sec.$IV$. At last in Sect. $V$, we finish up our outcomes.\\
\section{New Tsallis Agegraphic Dark Energy}
 For a flat FRW Universe filled with NTADE and   pressure less fluid, the first Friedmann equation is written as  
\begin{eqnarray}
\label{eq2}
H^2= \frac{1}{3m_{p}^2} {(\rho_{m}+\rho_{D})},
\end{eqnarray}
where $\rho_{D}$ is NTADE energy density  and $\rho_{m}$ is pressure less matter density. 
The fractional energy density is  easily instigate by  $ \Omega_{n} =\frac{\rho_{n}}{3m_{p}^2H^2}$ for $ n=~ m~$ and $~ D .$ Utilizing, this definition, the energy density parameter for matter and NTADE   are $\Omega_{m}=\frac{\rho_{m}}{3m_{p}^2H^2}$ and $
\Omega_{D}=\frac{\rho_{D}}{3m_{p}^2H^2}$,   respectively.
By putting the energy density parameter for matter and NTADE in  Eq. (\ref{eq2}), we will find  $\Omega_{m}=1-\Omega_{D}$ and, the energy densities ratio $\frac{\Omega_{m}}{\Omega_{D}}= r= \frac{1}{\Omega_{{D}}}-1$.\\

The conservation  law for matter  density and NTADE  is defined as :
\begin{eqnarray}
\label{eq7}
\dot{\rho}_{m} + 3H\rho_{m} = 0,
\end{eqnarray}
\begin{eqnarray}
\label{eq6}
\dot{\rho}_{D} + 3H\rho_{D}(1+\omega_{{D}}) =0.
\end{eqnarray}

In this model , the conformal is characterized as  $dt=ad\eta$ prompting  $\dot\eta=\frac{1}{ a}$ and accordingly

\begin{eqnarray}
\label{eq8}
\eta=\int_{0}^{a} \frac{da}{Ha^2},
\end{eqnarray}
and
\begin{eqnarray}
\label{eq9}
\rho_{D}= B \eta^{2\delta-4},
\end{eqnarray}
where $\eta=\Big({\frac {3{H}^{2} \Omega_{{D}}}{B}} \Big)^{\frac{1}{2\delta-4 }}$. Using conservation equations given in Eq. (\ref{eq7}) and  Eq. (\ref{eq6}), consolidating the  derivative of Eq. (\ref{eq2}) with time and  using Eq. (\ref{eq9}) and its time derivative,  we get :
\begin{eqnarray}
\label{eq10}
\frac{\dot H}{H^2}=-\frac{3}{2}(1-\Omega_{D})+\frac{(\delta-2)\Omega_{D}} {a \eta H},
\end{eqnarray}
by Eq.(\ref{eq10}), the deceleration parameter is  
\begin{eqnarray}
\label{eq13}
q=-1-\frac{\dot H}{H^2}=\frac{1}{2}-\frac{3}{2}\Omega_{{D}}-{\frac {(\delta-2)\Omega_{{D}}}{a \eta H}}.
\end{eqnarray}

By substituting Eq.(\ref{eq9}) and its time derivative into Eq.\ref{eq6} we find the EoS parameter as:
\begin{eqnarray}
\label{eq11}
\omega_{D}=-1- {\frac { 2\delta-4}{3 a \eta H}}.
\end{eqnarray}
\begin{figure}[htp]
	\begin{center}
		\includegraphics[width=9.3cm]{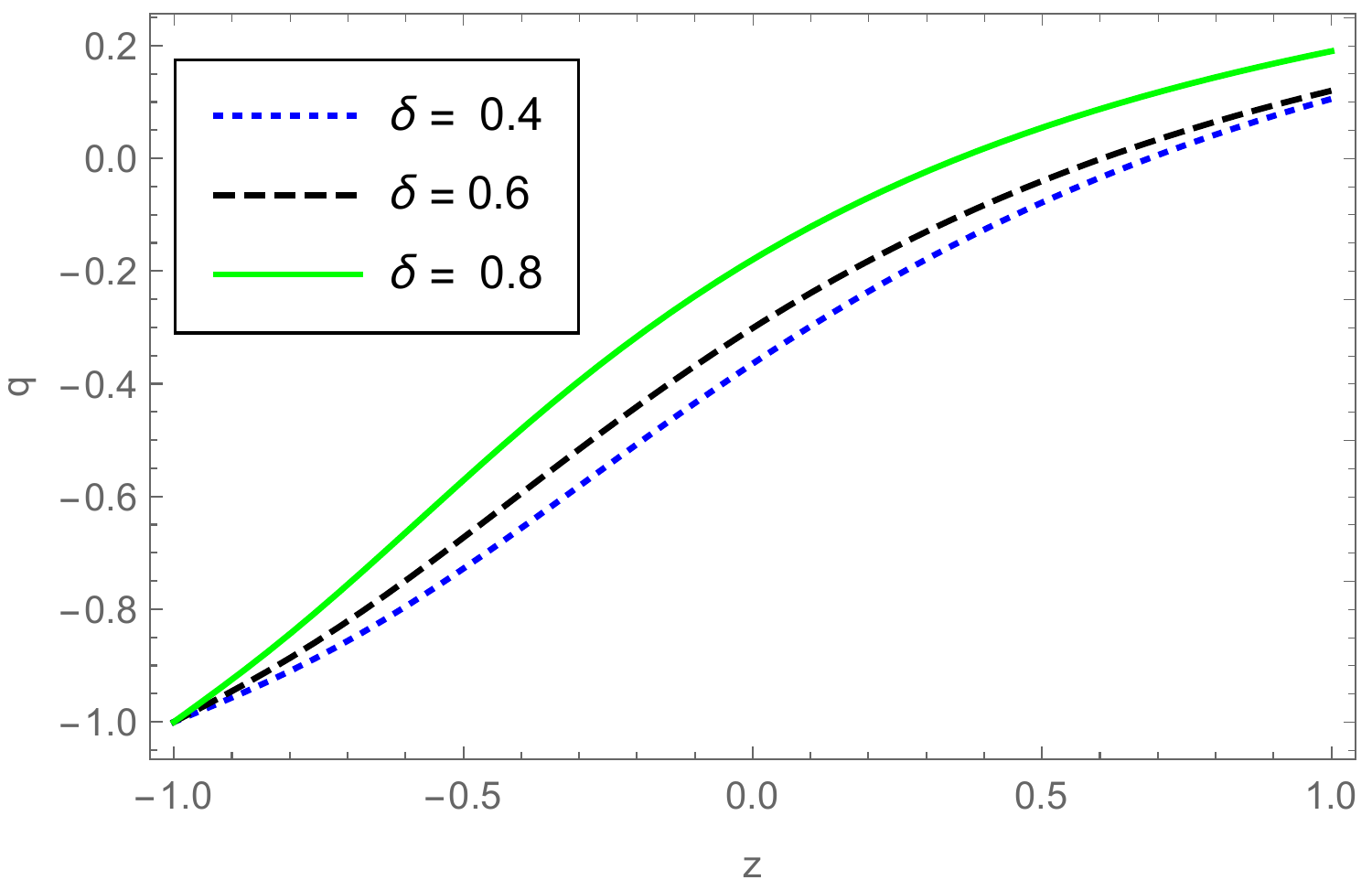}
		\includegraphics[width=9.3cm]{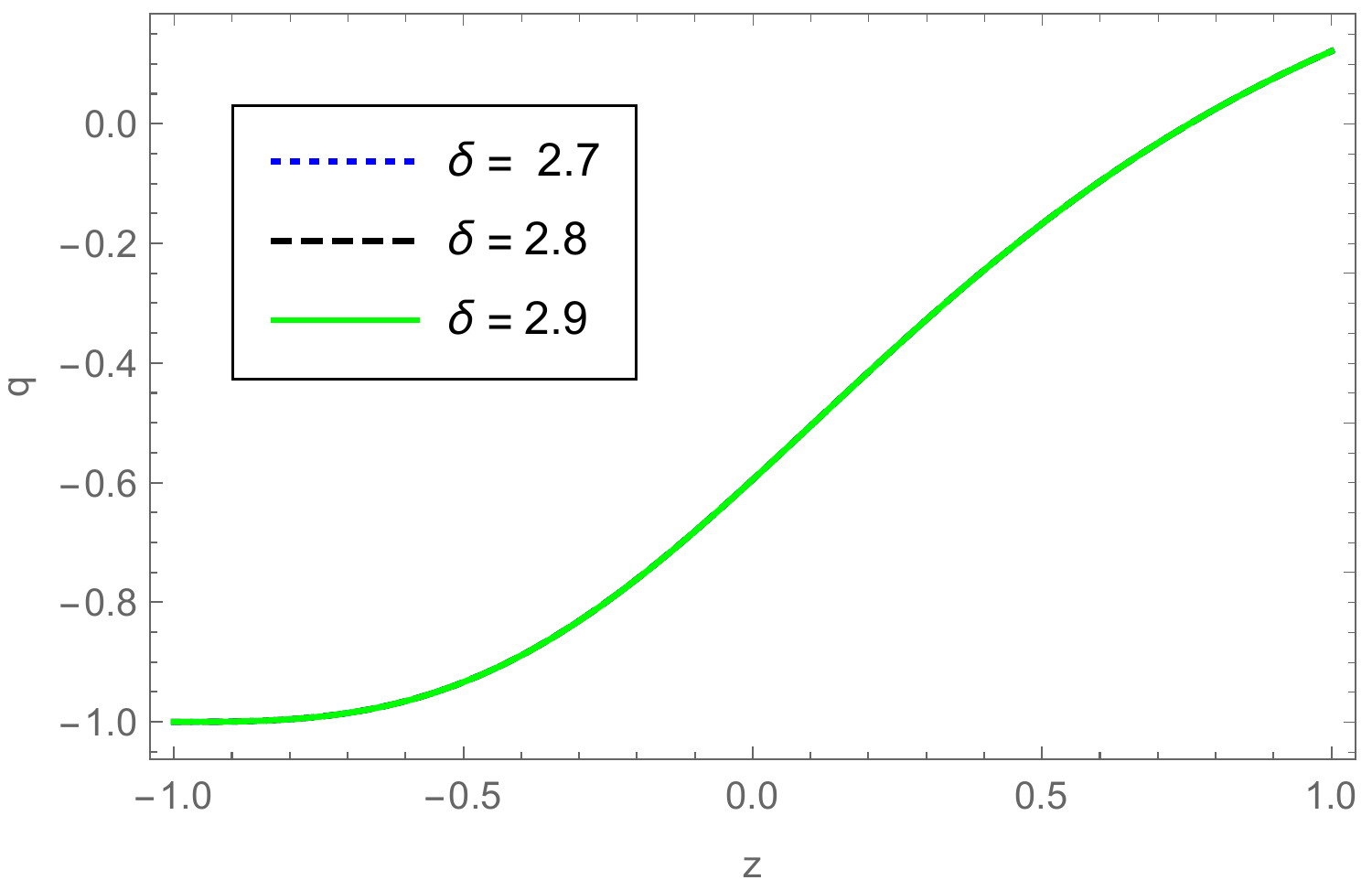}
	\caption{ Evolution of  deceleration parameter ($q$) against redshift parameter  ($z$) for without interaction NTADE. Here $\Omega_m=0.30$, $H(z=0)=67$and $B=3$}
		\label{qz non-interacting}
	\end{center}
\end{figure}

\begin{figure}[htp]
	\begin{center}
		\includegraphics[ width=9.3cm]{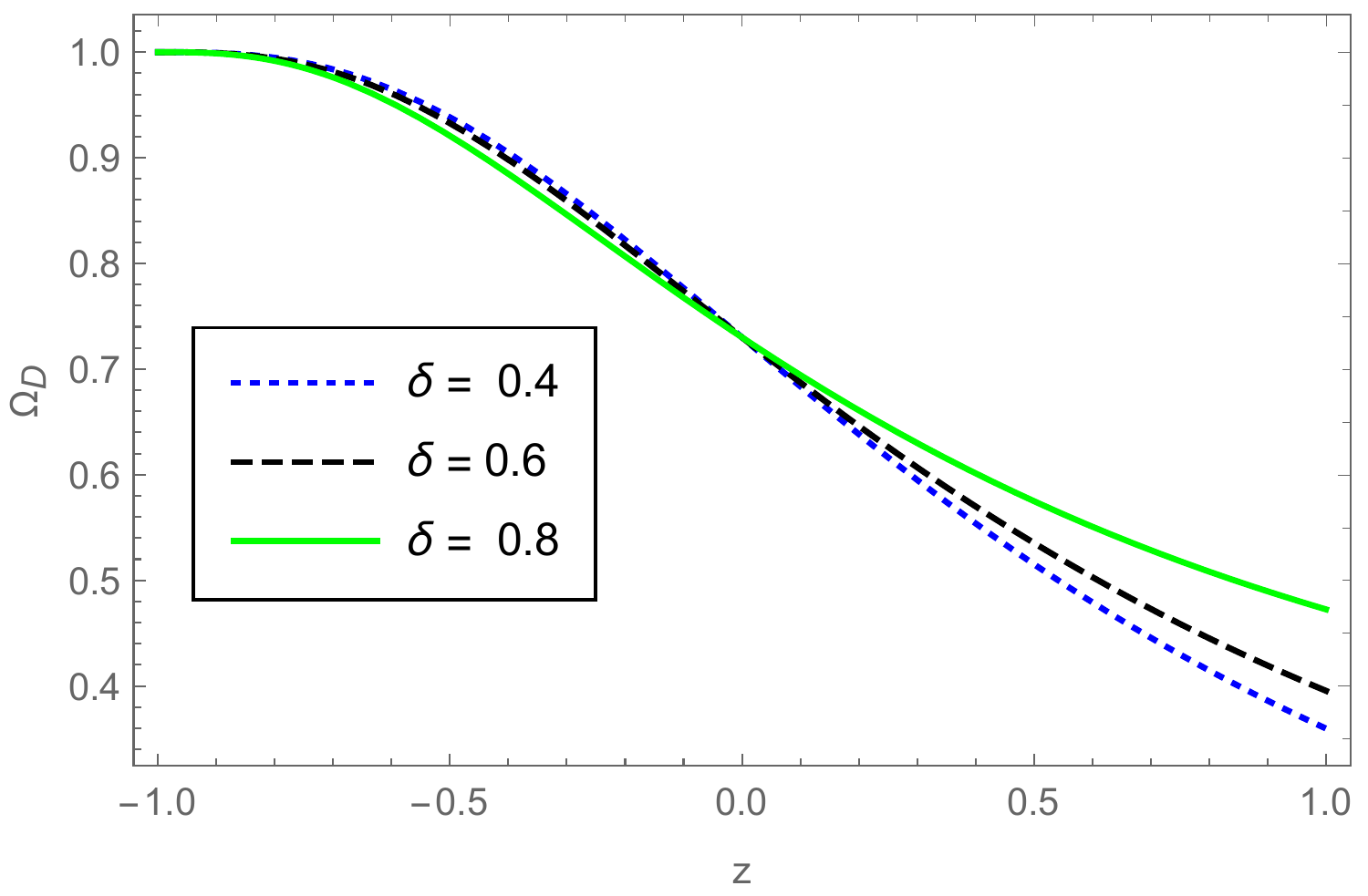}
		\includegraphics[ width=9.3cm]{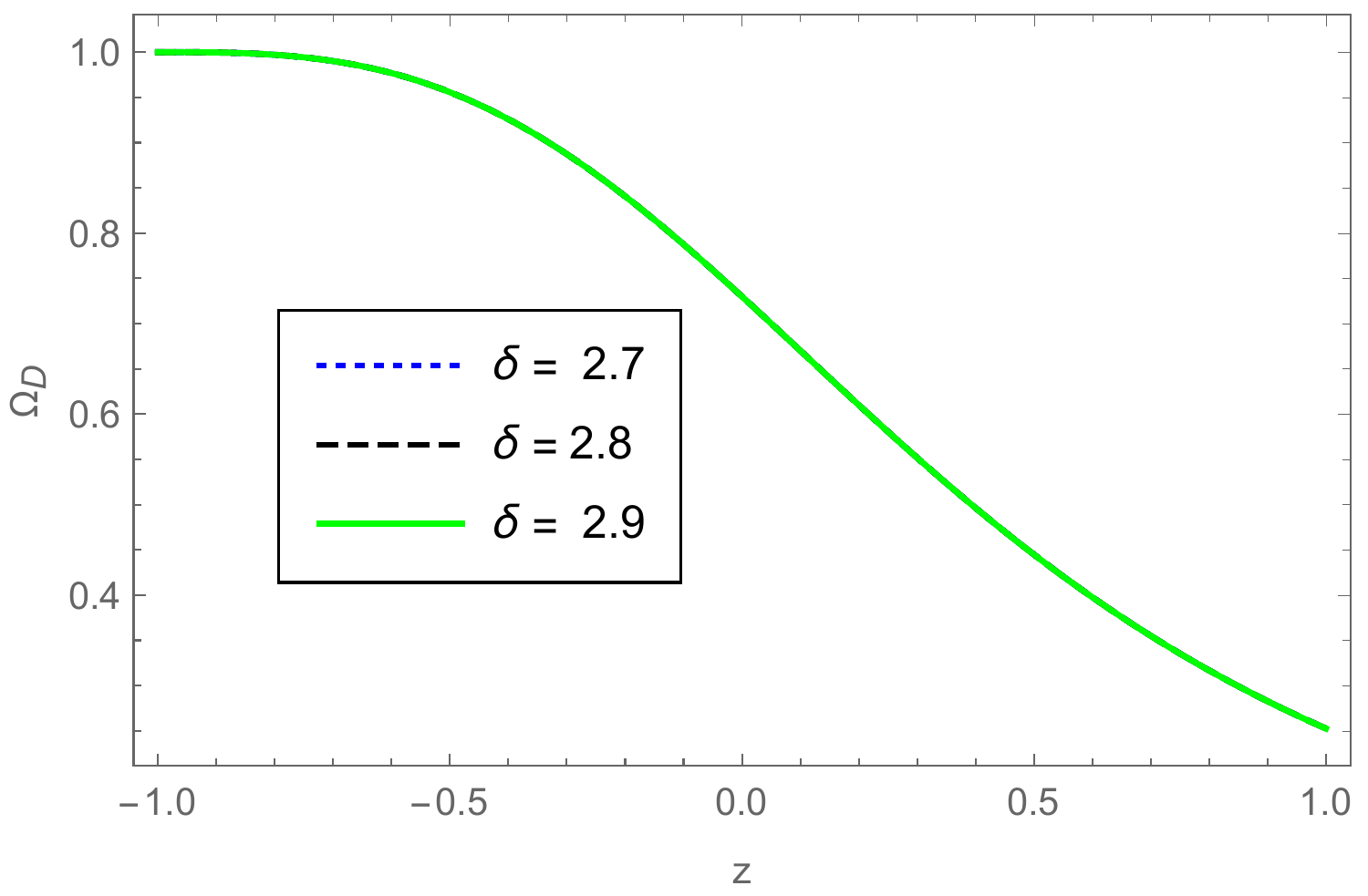}\\
		\includegraphics[  width=9.3cm]{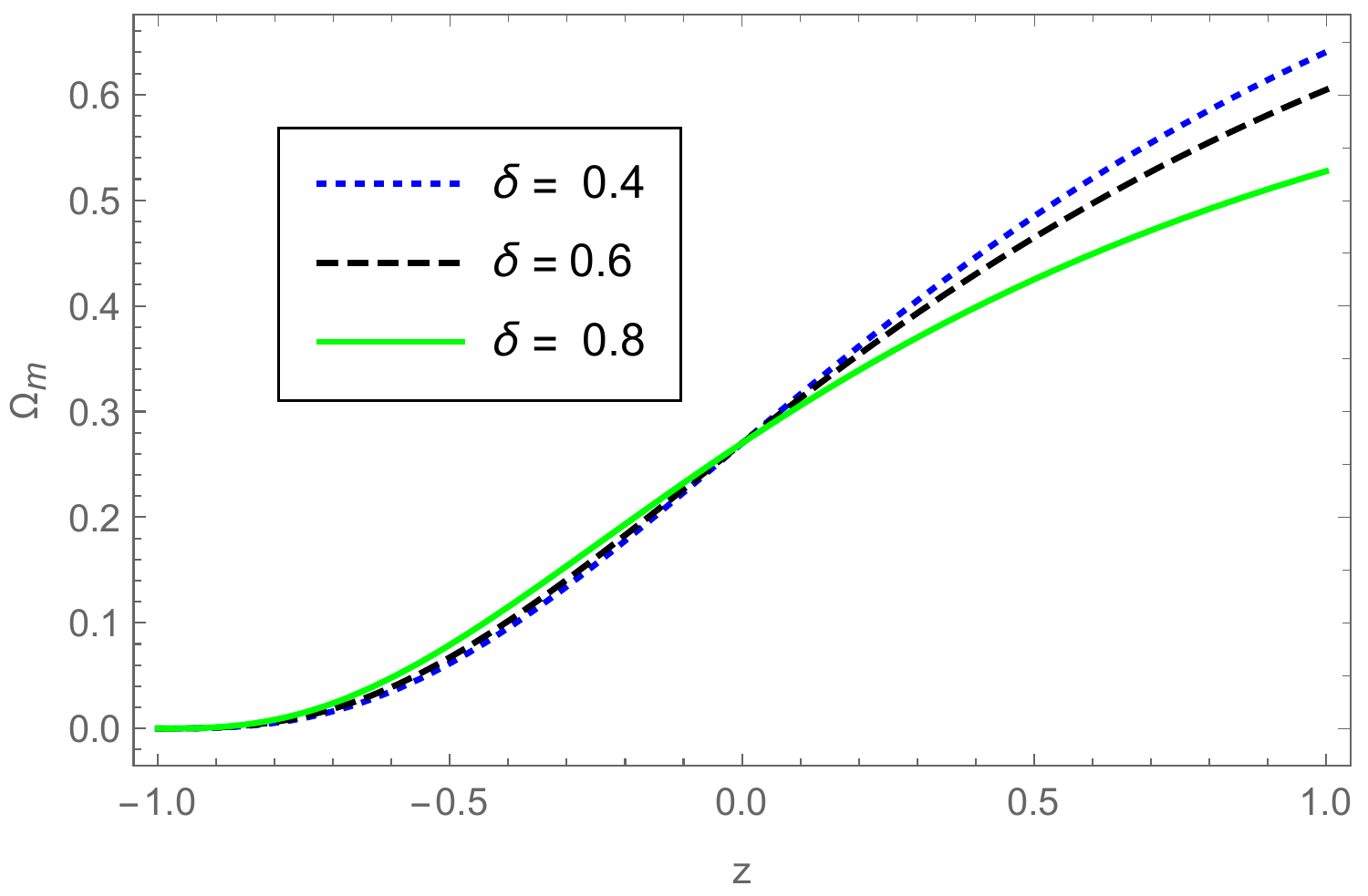}
			\includegraphics[ width=9.3cm]{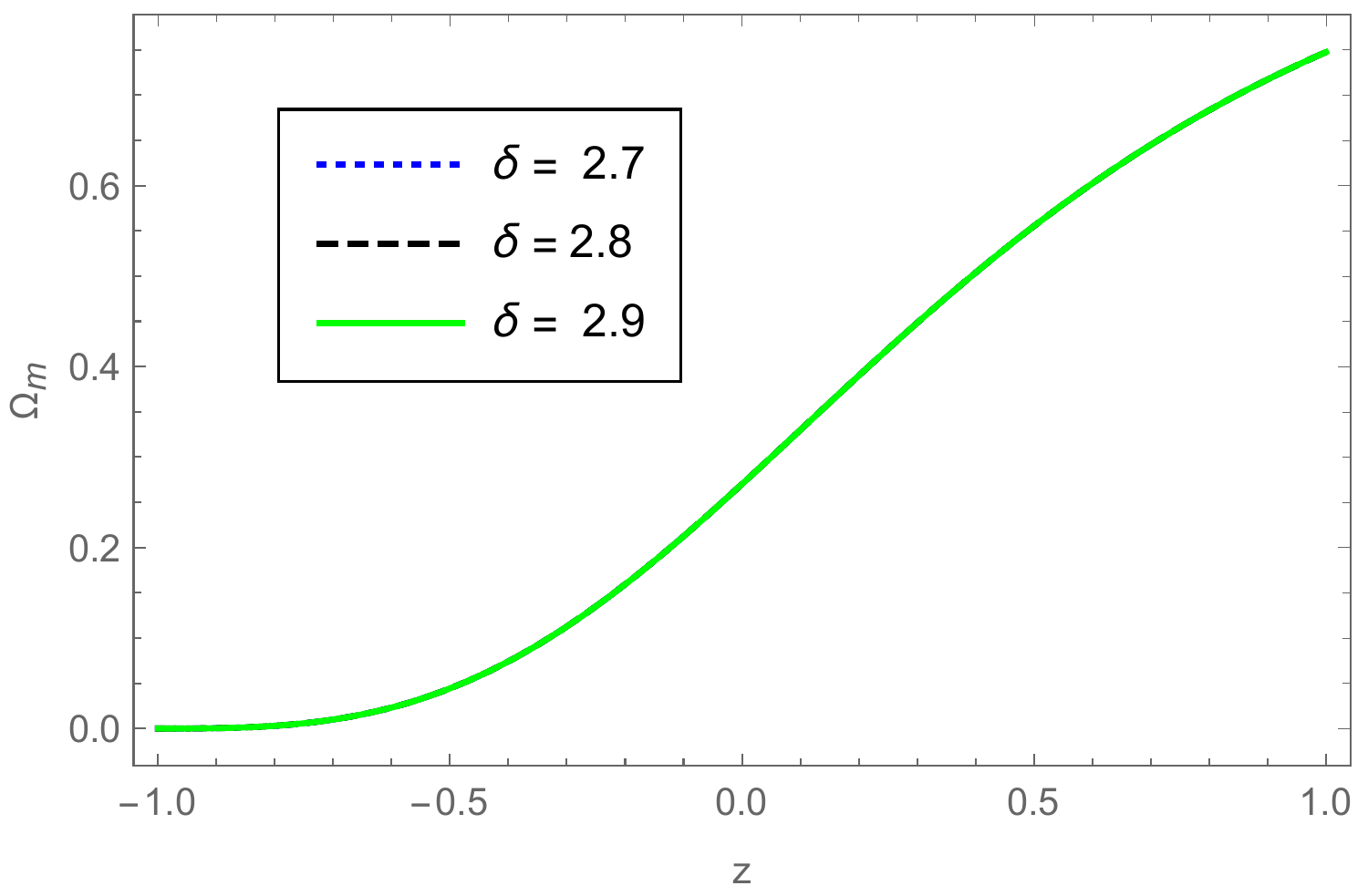}\\
		
	\caption{ The evolution of the dark energy density
	parameter 
	$\Omega_{D}$ and the matter density parameter $\Omega_{m}$,
	 as a function of the redshift z, for the scenario of modified
	cosmology through non-extensive thermodynamics with varying
	exponent. Here $\Omega_m=0.30$, $H(z=0)=67$ and $B=3$}
		\label{qz non-interacting}
	\end{center}
\end{figure}

\begin{figure}[htp]
	\begin{center}
		\includegraphics[width=9.3cm]{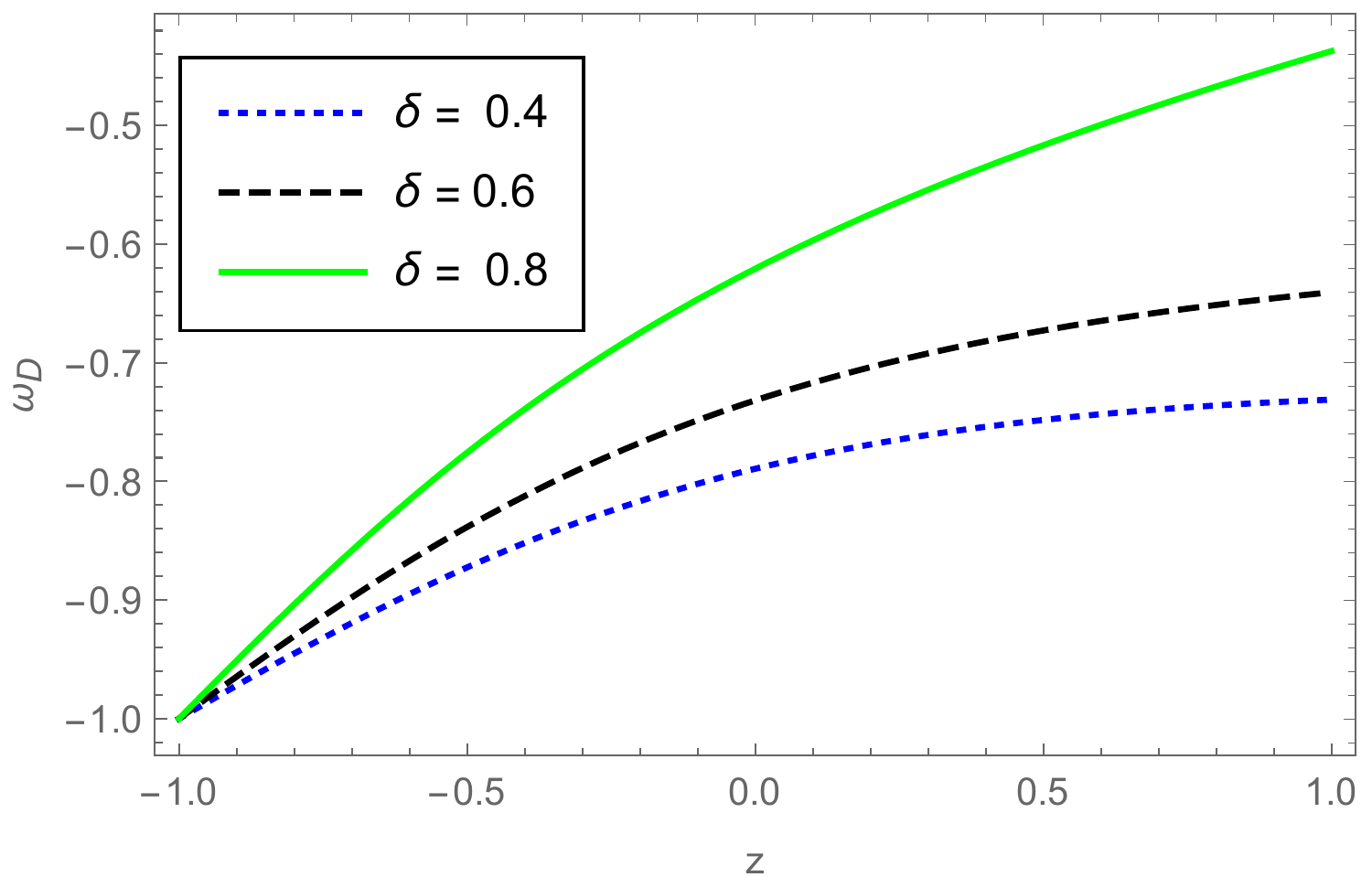}
		\includegraphics[width=9.3cm]{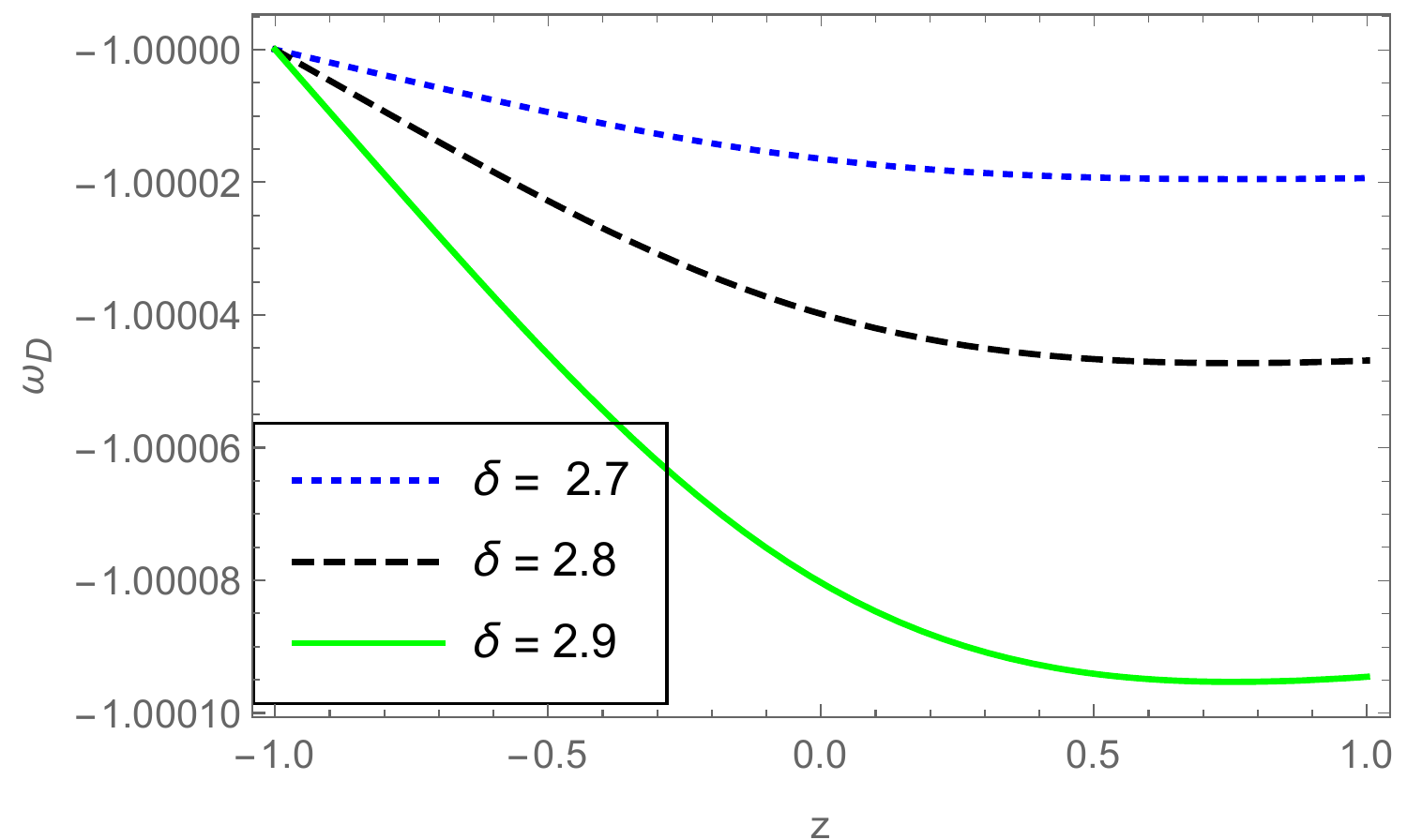}
			\caption{Evolution of  EoS parameter $\omega_D$ against  redshift parameter  ($z$) for without interaction NTADE. Here $\Omega_m=0.30$, $H(z=0)=67$and $B=3$}
		\label{fig$3$}
	\end{center}
\end{figure}

Using  Eq.(\ref{eq9}) and Eq.(\ref{eq10}), we get the derivatives of energy density parameter
\begin{eqnarray}
\label{eq14}
\dot\Omega_{D}=\frac {(2\delta-4) \Omega_{{D}}}{a\eta}+2\Omega_{{D}} H ( 1+q),
\end{eqnarray}
\begin{eqnarray}
\label{eq15}
\Omega_{D}^{'}=\frac {(2\delta-4) \Omega_{{D}}}{a \eta H}+2\Omega_{{D}}( 1+q).
\end{eqnarray}
Where, prime and dot  give the derivative with respect to $log a$  and time respectively.\\
By using Eq.(\ref{eq11}) and its time derivative we get the derivative  of EoS parameter with respect to $log a$

\begin{eqnarray}
\label{eq16}
\omega_{D}^{'}={\frac {(2\delta-4)}{3a \eta H}}\Big[ {\frac {1+ (\delta-2 ) \Omega_{{D}}}{a \eta H}}-\frac{3}{2}(1-\Omega_{{D}})\Big].
\end{eqnarray}
The evolutionary behavior of the deceleration parameter and EoS parameter is plotted for the NTADE model versus redshift $z$ by finding its numerical solution using the initial values of $\Omega_{m}$ as $\Omega_{m0}$= 0.30 and $H_{0}= 67$.  The deceleration or acceleration of the Universe is described by the deceleration parameter $q$. The Universe is in accelerating or decelerating phase accordingly if $q<0$ or $q >0 $,  respectively.  At present, it is seen that the Universe is accelerating with the goal that the estimation of the deceleration parameter lies in the range of $-1 \leq q < 0$. By Fig. 1,  we observe the behavior of the deceleration parameter against redshift ($z$).  It is  clear through  Fig. 1, that the NTADE  model transits from an early decelerating phase to a present accelerated phase for distinct values of the parameter $\delta$ in both  panels.\\

  The  modified cosmological scenario has been constructed by applying the 
	first law of thermodynamics and using the Tsallis entropy that possess the 
	usual one as a particular limit \cite{refG3}. Moreover,  Odinstov et al. \cite{ref38}, proposed this modified cosmological scenario with varying exponent $\delta$ by applying the non-extensive thermodynamics. Also, it is proposed that this modified cosmological scenario 
	may deliver a description of both inflation and late-time acceleration of the Universe with  varying-exponent from the first law of 
	thermodynamics \cite{ref38}.\\

	 In Fig. 2 we depict the energy density for NTADE $\Omega_{D}$ and energy density of matter $\Omega_{m}$ as a
	function of redshift. From this figure, we can see that we obtain
	the usual thermal history of the universe, namely the successive
	sequence of matter and dark-energy epochs. It is also proposed by researchers that concerning the universe evolution at late times, the new
	terms that appear due to the non-extensive varying exponent
	constitute an effective dark energy sector \cite{ref38,ref45,refG4,refG5}. As we showed, the
	universe exhibits the usual thermal history, with the successive
	sequence of matter and dark-energy epochs, and with
	the transition to acceleration  in
	agreement with the observed behavior.\\

	 The EoS parameter $\omega_{D}$ with respect to $z$ is plotted for different parameter $\delta$  in Fig. 3. From the upper panel of Fig. 3, we observe that the NTADE EoS parameter  $\omega_{D}$ lies in the quintessence region, not crosses the phantom divide line  and finally approaches the CC i.e. ($\omega_{D} = -1$) at the future for $\delta$ =0.4, 0.6 and 0.8.  From the lower panel of Fig. 3, it is also clear that the NTADE EoS parameter  $\omega_{D}$ lies in the phantom region, not crosses the phantom divide line and finally approaches the CC i.e. ($\omega_{D} = -1$)  in the future for $\delta$ =2.7, 2.8 and 2.9.   It is worth noting that the NTADE EoS parameter gives a nice behavior and it can be quintessence-like and phantom-like depending upon the different values of $\delta$. The similar behavior of the deceleration and the EoS parameter   has also been observed in  \cite{Zadeh:2018wub}\\

\section{NTADE : Statefinder Analysis}
In this section, we discuss  the NTADE model  without interaction  through the statefinder and snap parameter analysis. From the definitions of Eq. (\ref{eq1}), we may write

\begin{eqnarray}
\label{eq23}
r=1+\frac{9}{2}\Omega_{{D}}~\omega_{{D}} (1+\omega_{{D}})-\frac{3}{2}\omega_{D}^{'}~\Omega_{{D}},
\end{eqnarray}
\begin{eqnarray}
\label{eq24}
s=(1+\omega_{{D}})-{\frac {\omega_{D}^{'}}{3\omega_{{D}}}}.
\end{eqnarray}
 In cosmology, the snap parameter is defined as \cite{refG6}
\begin{eqnarray}
\label{eq24}
 A_{4}=\frac{\ddddot{a\hspace{0pt}}}{aH^{4}}.
\end{eqnarray}

Snap (the fourth order time
derivative) is also sometimes called jounce \cite{refG6}. This parameter appears in the fourth order
term of the Taylor expansion of the scale factor around $a_{0}$, where the subscript 0 denotes the present-day value.  The snap  parameter $A_{4}$ can be expressed as
elementary function of the 
the density parameter $\Omega_{m}$ \cite{refG7}.

\begin{eqnarray}
\label{eq24}
 A_{4}=1-\frac{9}{2}\Omega_{m}.
\end{eqnarray}
The different DE models can be discriminated  against each other and from the $\Lambda$CDM model using  first statefinder  parameters $r$,  second statefinder  parameter $s$, ($ r, s$) plane and ($ r, q$) plane.  Several dark energy models  show distinct evolutionary trajectories of their evolution in  ($ r, s$) and ($ r, q$) plane. Now, we use $r$  (first statefinder  parameter) , $s$ (second statefinder  parameter) , ($ r, s$) plane and ($ r, q$) plane to discriminate the NTADE model for different  values of the parameter $\delta$, while fixing other parameters according
	to the best-fit observational values. Also, an important purpose of any diagnostic is that it allows us to discriminate between a given DE model and the simpler of all models -  $\Lambda$CDM. This is exactly done by the  statefinder.  The value of  first statefinder parameter $r$ remains pegged at unity i.e. $r= 1$ for the $\Lambda$CDM model, even as the matter density evolves from a large initial value to a small late-time value . It is easy to show that $({r, s})=({1, 0})$ is a fixed point for $\Lambda$CDM \cite{Sahni:2002fz,Alam:2003sc}. \\


\begin{figure}[htp]
	\begin{center}
		\includegraphics[width=7cm,height=7.5cm]{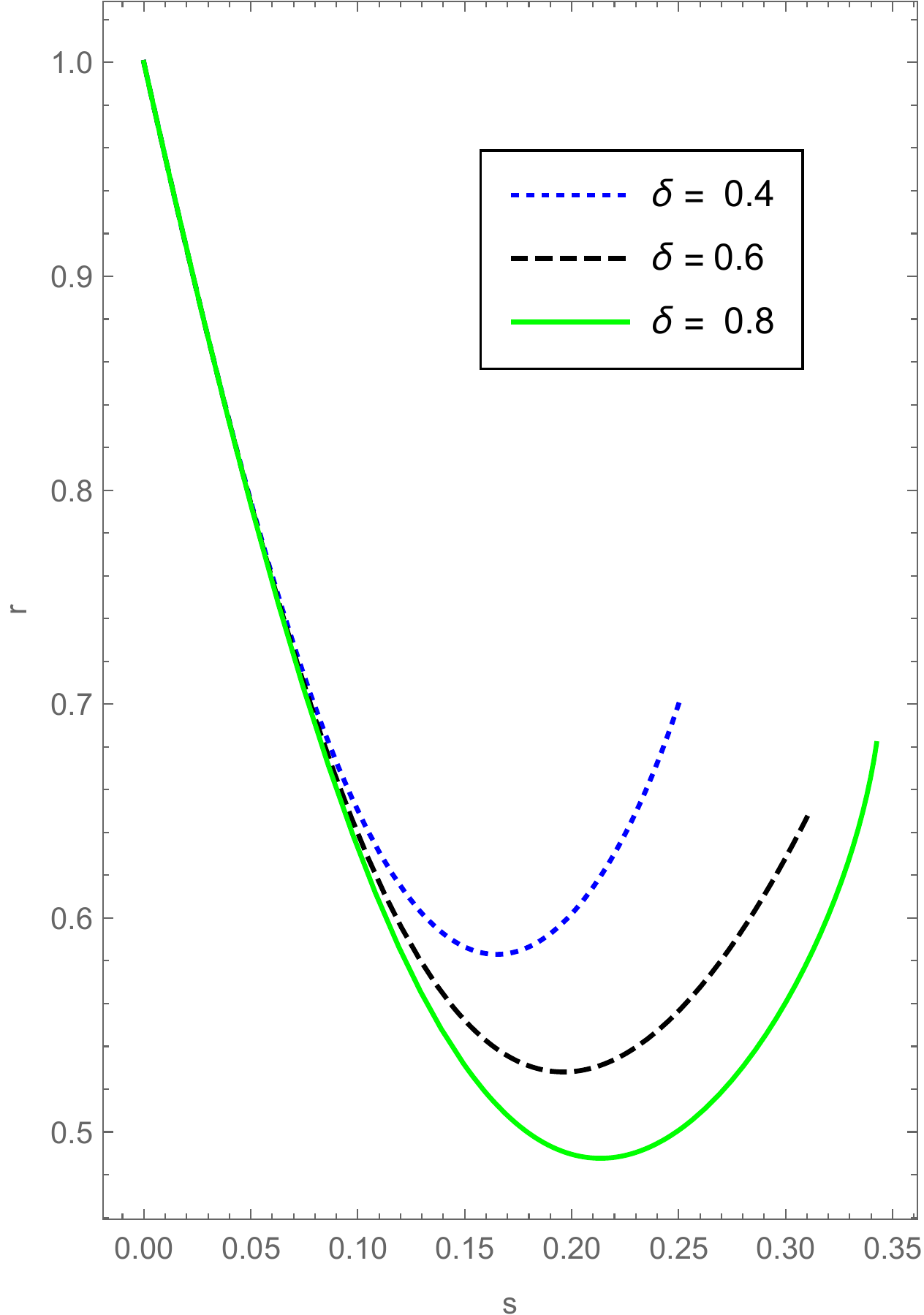}
		\includegraphics[width=7cm,height=7.5cm]{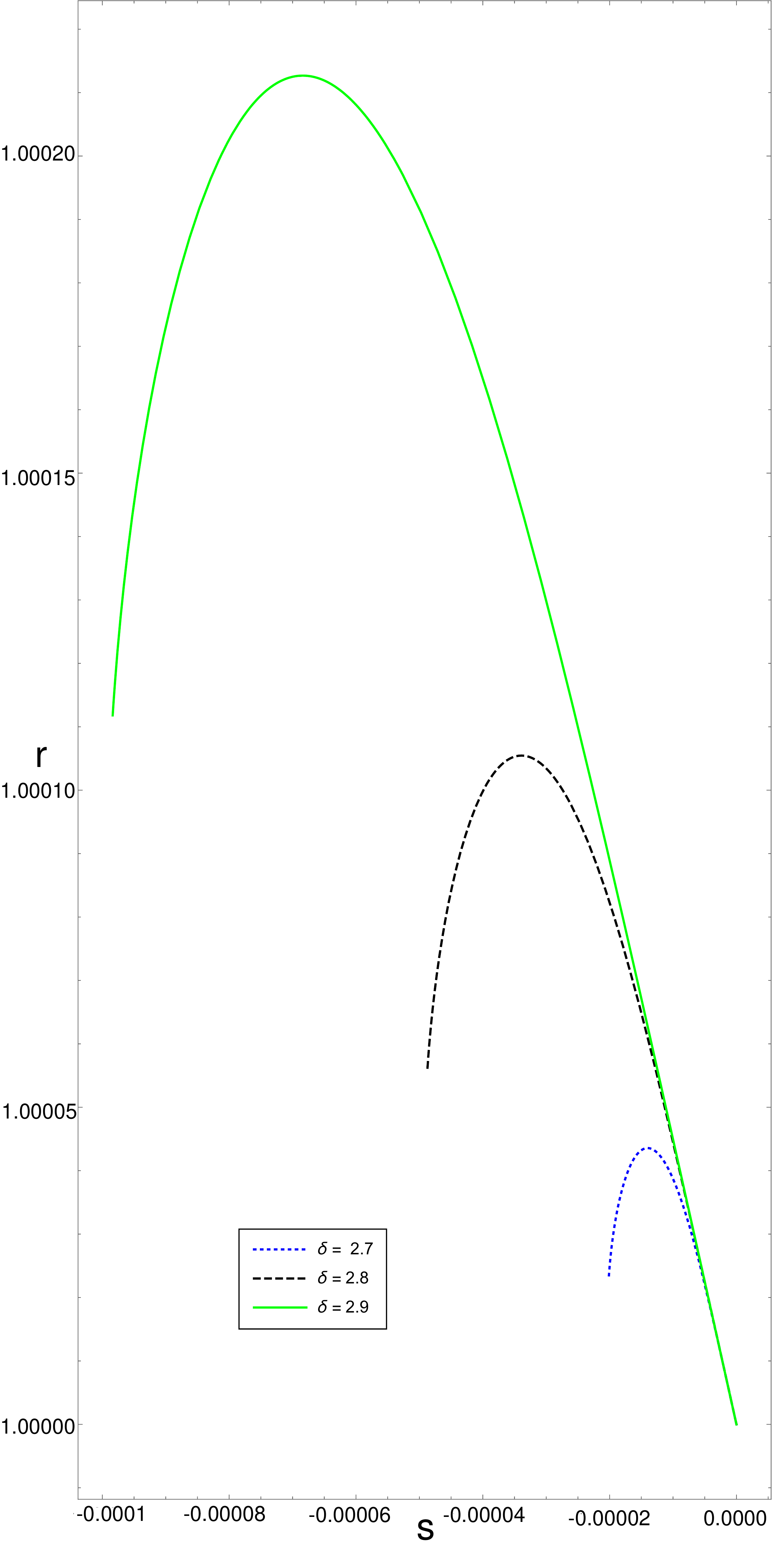}
	\caption{Evolution trajectories of the statefinder in the $r-s$ for different $\delta$. Here $\Omega_m=0.30$, $H(z=0)=67$ and $B=3$}
	\label{ rs non-interacting}
	\end{center}
\end{figure}

\begin{figure}[htp]
	\begin{center}
		\includegraphics[width=7cm,height=7.5cm]{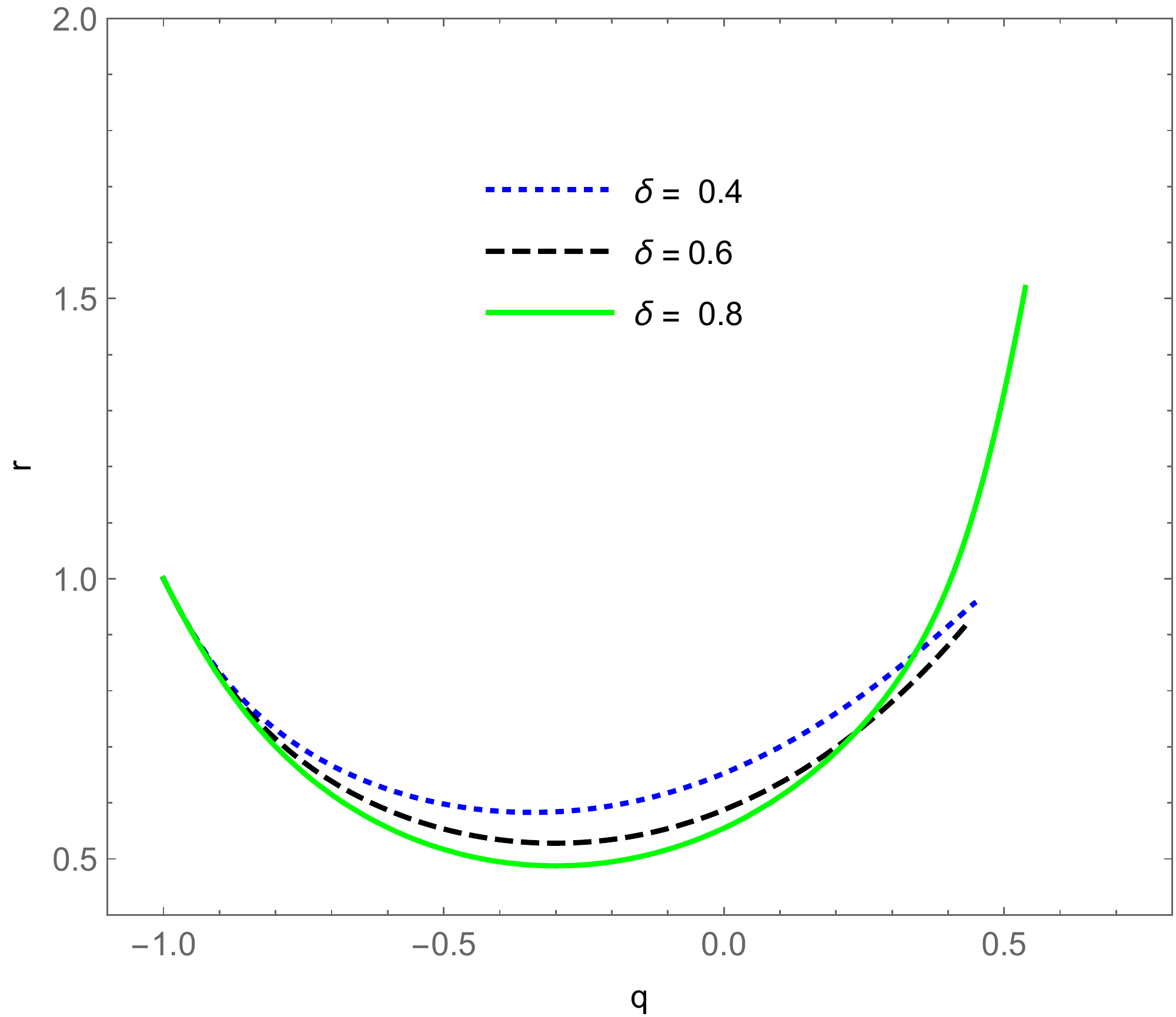}
		\includegraphics[width=7cm,height=7.5cm]{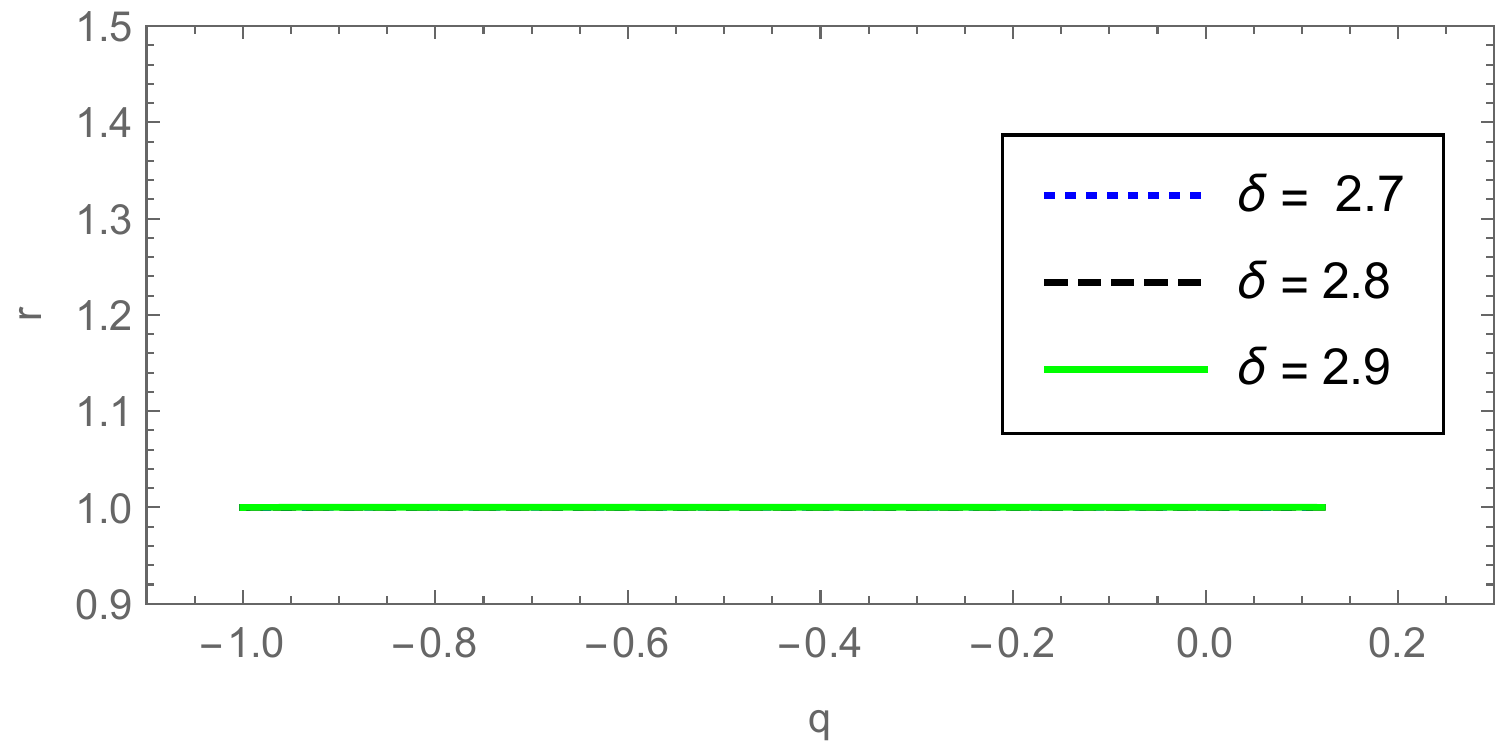}
		\caption{Evolution of  first statefinder parameter  ($r$ )  against the  deceleration parameter ($q$)  without interaction NTADE. Here $\Omega_m=0.30$, $H(z=0)=67$ and $B=3$}
	\label{rq non-interacting}
\end{center}
\end{figure}

\begin{figure}[htp]
	\begin{center}
		\includegraphics[width=7cm,height=7.5cm]{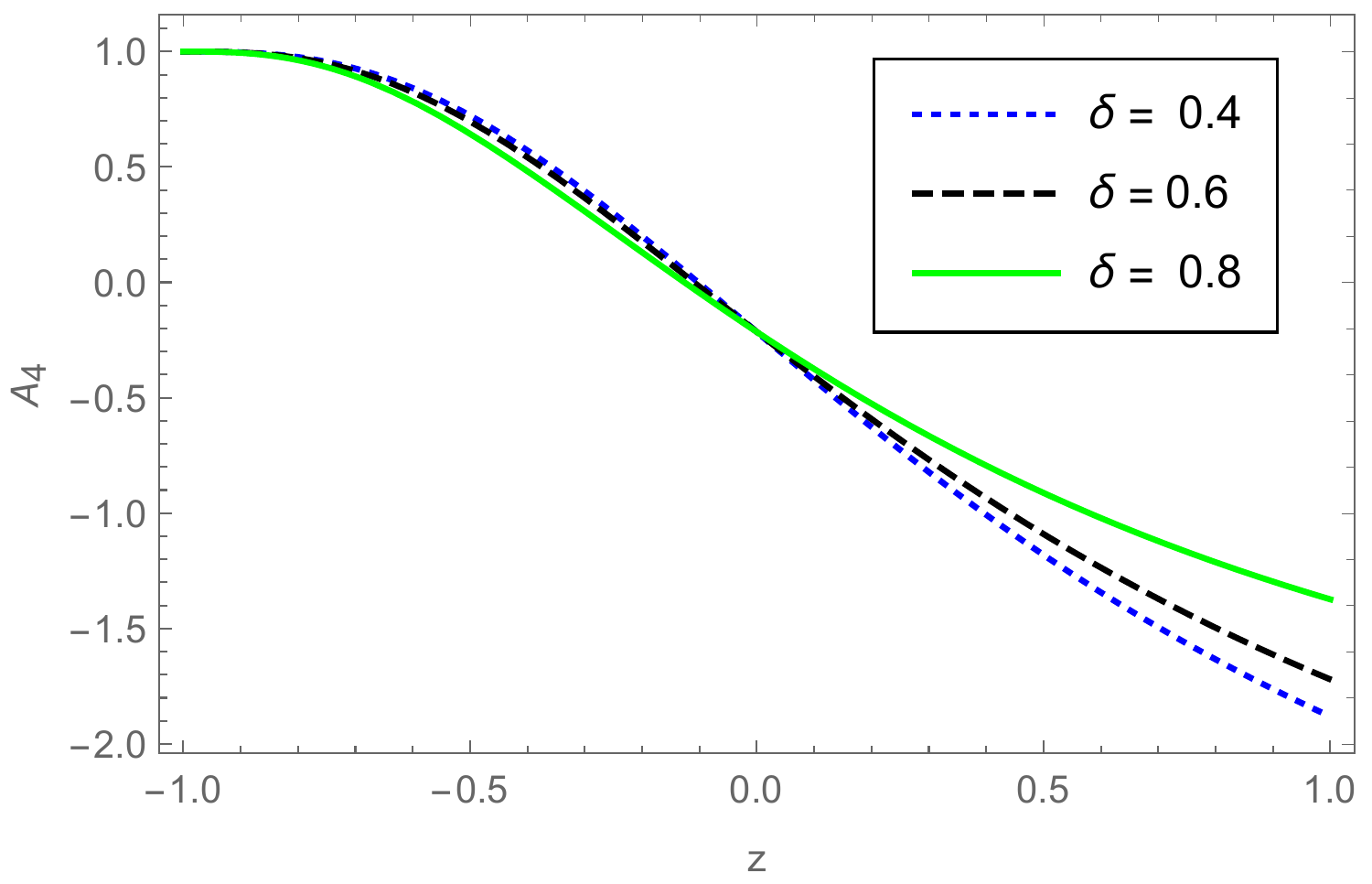}
		\includegraphics[width=7cm,height=7.5cm]{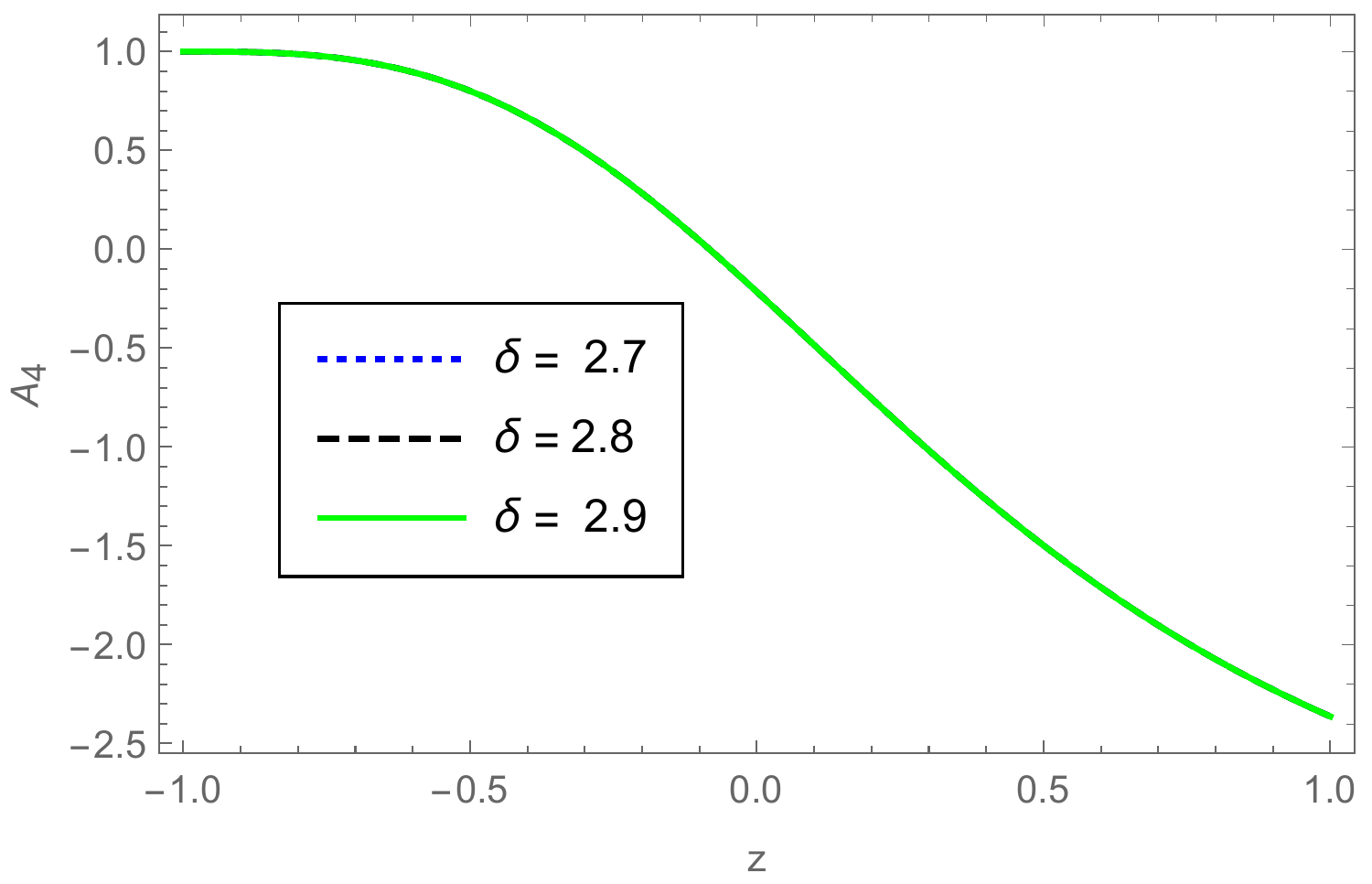}
		\caption{Evolution of the snap parameter $A_{4}$ with respect to redshift $z$. Here $\Omega_m=0.30$, $H(z=0)=67$ and $B=3$.   }
	\label{rq non-interacting}
\end{center}
\end{figure}

The evolutionary trajectory of the statefinder pair $(s, r)$ for the NTADE model is graphed in Fig. 4, for both the panels. From the upper panel of Fig. 4, we observe that the statefinder evolution trajectories   start with near the point $ \{r =0.7,s = 034\} $ and ends  at the $\Lambda$CDM fixed point ($r = 1$, $s = 0$). The statefinder analysis  has also been performed for the HDE, THDE, ADE and RDE models by the authors in \cite{ref56,ref57,ref45,ref56a}.   The HDE, ADE  and  the THDE models have shown  quite similar  behavior in the $ s - r $ plane to the NTADE model without interaction, while the NTADE model  has different evolutionary  behavior as compared to the  RDE model in the $ s - r $ plane \cite {ref54}. The evolutionary trajectories of the lower panel of   Fig. 4, evolve above the $r = 1$ and finally approaches to the $\Lambda$CDM fixed point.\\

In Fig. 5, we have plotted the evolutionary trajectories of the statefinder pair ($q-r$) for the NTADE model for both the panels. The fixed point $(q = 0.5, r = 1)$ presents the SCDM i.e. the matter dominated and (q = -1, r = 1) presents
the SS model i.e. de-Sitter universe, respectively.  The evolutionary trajectories of the ($q-r$) plane of the NTADE model starts from  the matter-dominated universe i.e. SCDM ( r = 1, q = 0.5) in the past and decreases monotonically and finally reaches to the de-Sitter expansion (SS) $(q = -1, r = 1)$ for different values of $\delta$ (upper panel). The evolutionary trajectories of the $(q, r)$ plane for $\delta=0.8$ starts with  another fixed point (upper panel). While  the evolutionary trajectories of the lower panel are horizontal line segment approaching towards the de-Sitter expansion (SS) $(q = -1, r = 1)$ for different values of $\delta$.\\

 Expanding the scale factor to fourth order with respect to time  is
physically meaningful, since cosmological data already allow to measure the third-order
term: the jerk parameter \cite{refG8}. The snap parameter may be important for observations
involving redshifts $z \sim 1$. and higher, where  the expansion of cosmological quantities in
powers of $z$ cannot be limited only to linear and quadratic terms \cite{refG9}. In  \cite{refG7}, the authors proposed that $A_{n}$ parameters evolve with time with  odd $A_{2n+1}$ (even $A_{2n}$) remaining larger (smaller) than unity. All
$A_{n}$ parameters approach unity in the distant future. We have plotted the snap parameter $A_{4}$ against redshift $z$ in Fig. 6 for different $\delta$. It can be seen from Fig. 6, $A_{4}$ evolves from past to present and increases monotonically at the  future. It is interesting to note that  the snap parameter $A_{4}$ remains smaller than unity and approaches unity in the distant future: $A_{4} \rightarrow 1$ when $z \rightarrow -1$.

\section{Analysis of the $\omega_{D}-\omega_{D}^{'}$ pair}
In this section, we study the $\omega_{D}- \omega_{D}^{'}$ pair dynamical analysis for the NTADE model which is widely used in the previous works.  The fixed point $\omega_{D} = -1$, $\omega_{D}^{'}=0$ denotes the standard $\Lambda$CDM  in the $\omega_{D}-\omega_{D}^{'}$ plot. The scopes of the DE quintessence model  have been explored in \cite{ref37a} in the $\omega_{D}-\omega_{D}^{'}$ plane.  The dynamical property of other dark energy models have been utilized from $\omega_{D}-\omega_{D}^{'}$  viewpoint \cite{ref79,ref80,ref81}. \begin{figure}[htp]
	\begin{center}
		\includegraphics[width=7cm,height=7.5cm]{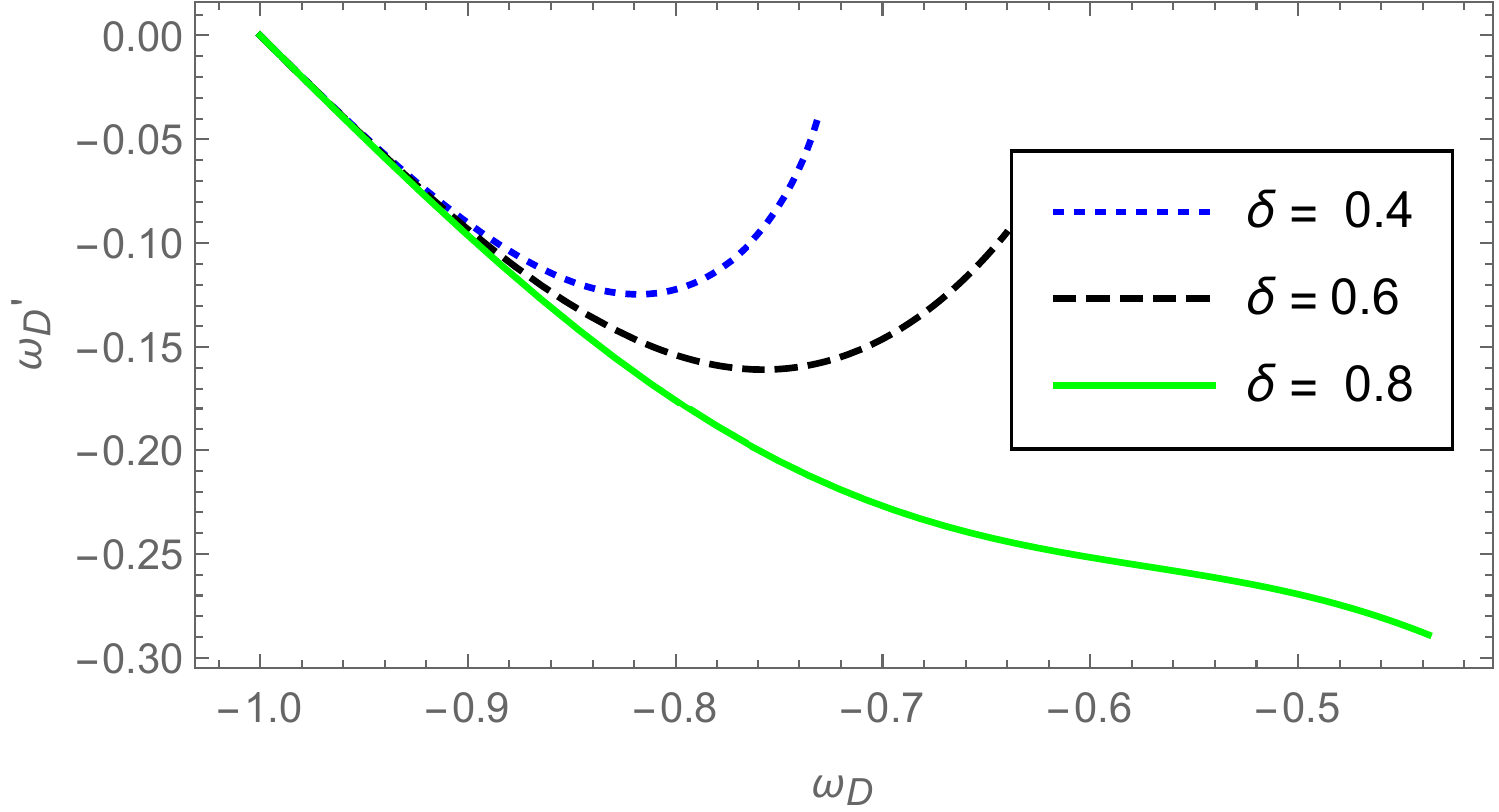}
		\includegraphics[width=7cm,height=7.5cm]{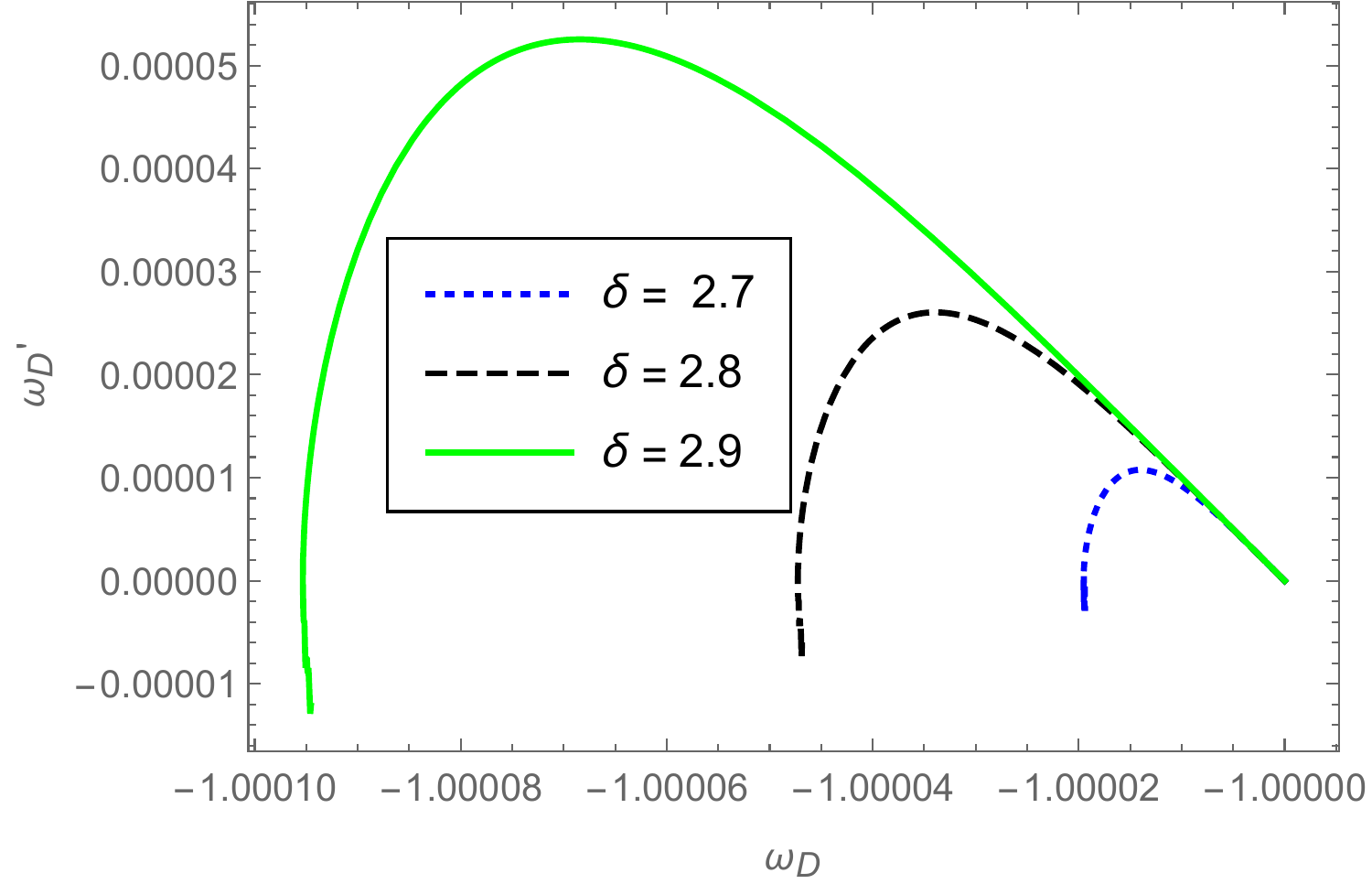}
\caption{ The  diagram of $\omega_{D}-\omega_{D}^{'}$  for  without interaction NTADE  for different model parameters $\delta$. Here $\Omega_m=0.30$, $H(z=0)=67$ and $B=3$}
\label{fig4}
	\end{center}
\end{figure}

The evolutionary trajectories of $\omega_{D}^{'}$ and $\omega_{D}$ plane are plotted in Fig. 7 for the NTDAE model for different values of $\delta$ in the upper and lower panel.  From the upper panel of Fig. 7, we observe that  the evolutionary trajectories vary in quintessence $\omega >-1$ and finally attains to $\Lambda$CDM ($\omega =-1$). while in  the lower panel of Fig. 7,  the evolutionary trajectories vary in the phantom region $\omega >-1$ and finally attains to $\Lambda$CDM ($\omega =-1$). In both the panels, the phantom divide line is not crossed.
\section{Closing remarks} 
In the present work, we have discuses New Tsallis Agegraphic dark energy model without interaction  by taking various values of non-extensive parameter $\delta$ within  framework of 
 a flat FRW Universe. The main objective of this article is to distinguish the NTADE model from the $\Lambda$CDM model through  statefinder diagnostic. We can summarize our results as:
\begin{itemize}  	
	
	\item We observe from the evolutionary behavior of  the deceleration parameter $q$ that, it
	exhibits a transition from the early decelerated phase to late time accelerated phase. 
	
	\item It can be  observed  that the NTADE EoS parameter shows a rich behaviour, it can be 
		quintessence-like or phantom-like depending on the value of $\delta$.
	
	\item  The evolutionary trajectory in $(s, r)$ plane of the NTADE model shows that it approaches 
	the  $\Lambda$CDM  fixed point $(r = 1, s = 0)$ at   the future.

	\item The evolutionary behavior of ($q, r$) plane of the NTADE model indicates that it evolves  from the
	matter dominated Universe i.e. SCDM $( r = 1, q = 0.5)$  in early time and approaches to the point $(q = -1, r = 1)$ i.e. the de Sitter expansion (SS) in the far future for various values of the parameter $\delta$.
	
	\item We observe from the dynamical analysis $\omega_{D}- \omega_{D}^{'}$ pair for the NTADE model that the evolution trajectories lie either in the quintessence region or phantom region depending on the value of the parameter  $\delta$.\\
	
	The statefinder analysis  has also been performed for the HDE, THDE, ADE and RDE models  in literature as mentioned in   the introduction.  The HDE, ADE  and  the THDE models have shown   quite similar behavior in the $ s - r $ plane to the NTADE model without interaction, while the NTADE model  has different evolutionary behavior as compared to the  RDE model in the $ s - r $ plane.\\
	
	In the   future paper, we shall investigate the effects of the interaction between DE and DM  on the NTADE model through the statefinder. Also, other diagnostics such as Om diagnostic can also be studied to understand the nature of the NTADE model.  
	
\end{itemize}

\acknowledgments{The authors  gratefully acknowledge the use of services and facilities provided by  GLA University, Mathura, India to conduct this research work.}  We would like to thank the learned reviewer(s) for his valuable comments and suggestions that help us to improve this paper in the present form.

\end{document}